\documentclass{ws-ijmpa}
\usepackage[super,compress]{cite}
\usepackage{graphicx}
\usepackage{epsfig}
\usepackage{amssymb}
\usepackage{graphicx}% Include figure files
\usepackage{dcolumn}% Align table columns on decimal point
\usepackage{bm}% bold math

\begin{document}

\markboth{Yupeng Yang}
{Constraints on the primordial power spectrum of small scales using the neutrino signals from the dark matter decay}

\title{Constraints on the primordial power spectrum of small scales using the neutrino signals from the dark matter decay}
\author{Yupeng Yang}%$^{1,2}$\thanks{E-mail:aynuyyp@163.com}\\
\address{Collage of Physics and Electrical Engineering, Anyang Normal University, \\Anyang, 455000, China
\\Joint Center for Particle, Nuclear Physics and Cosmology, \\Nanjing, 210093, China\\
aynuyyp@163.com}

%\date{Accepted 1988 December 15. Received 1988 December 14; in original form 1988 October 11}

\maketitle

\begin{abstract}
Many inflation theories predict that the primordial power spectrum is scale invariant. 
The amplitude of the power spectrum can be constrained by different 
observations such as the cosmic microwave background (CMB), Lyman-$\alpha$, large-scale structures 
and primordial black holes (PBHs). Although the constraints from the CMB are 
robust, the corresponding scales are very large ($10^{-4}<k<1 \mathrm{Mpc^{-1}}$). 
For small scales ($k > 1 \mathrm{Mpc^{-1}}$), the 
research on the PBHs provides much weaker limits. Recently, 
ultracompact dark matter minihalos (UCMHs) was proposed and it was found that they 
could be used to constraint the small-scale primordial power spectrum. 
The limits obtained by the research on the UCMHs are much better than that of PBHs. 
Most of previous works focus on the dark matter annihilation within the UCMHs, but 
if the dark matter particles do not annihilate the decay is another important issue. 
In previous work~\cite{EPL}, we investigated the gamma-ray flux 
from the UCMHs due to the dark matter decay. In addition to these flux, the neutrinos 
are usually produced going with the gamma-ray photons especially 
for the lepton channels. In this work, we studied the neutrino 
flux from the UCMHs due to the dark matter decay. Finally, we got the constraints 
on the amplitude of primordial power spectrum of small scales.

\end{abstract}

\ccode{PACS numbers: 95.35.+d}

\section{Introduction}           %% first-level sections will be auto-capitalized
\label{sect:intro}
It is well known that the cosmic structures are originated from the primordial 
density perturbations which are produced during the inflation. 
Most of inflation theories predict that the primordial power spectrum (PPS) 
of density perturbations is scale invariant~\cite{Lidsey:1995np}. The PPS ($\mathcal{P}_\mathcal{R}(k)$)
\footnote{In this paper, we use the curvature perturbations ($\mathcal{P}_\mathcal{R}$) 
instead of the density perturbations ($\mathcal{P}_{\delta}$). 
In the picture of the scale invariant, there is a very simple linear relation 
between of them~\cite{Bringmann:2011ut}.}
can be constrained 
by some observations. At present, the limits are mainly from the 
observations on the cosmic microwave background (CMB), Large-scale structures, 
Lyman-$\alpha$ forest and microlensing effect~\cite{Larson:2010gs,Hlozek:2011pc,Bird:2010mp,Tinker:2011pv}. 
But these constraints focus on large scales, $k - 10^{-4} \sim 1 \mathrm{Mpc^{-1}}$. 
For small scales, $k > 1 \mathrm{Mpc^{-1}}$, 
the limits are mainly from the research on the primordial black holes (PBHs)~\cite{Josan:2009qn}. 
However, the constraints from the PBHs are about 7 orders weaker 
($\mathcal{P}_\mathcal{R}(k) \sim 10^{-2}$) 
than that from the CMB ($\mathcal{P}_\mathcal{R}(k) \sim 10^{-9}$). 
Recently, one new kind of dark matter structures named 
ultracompact dark matter minihalos (UCMHs) was proposed and they 
could be used to constrain the PPS of small-scales~\cite{Ricotti:2009bs}. 
Up to date most of works focus on the dark matter annihilation in the UCMHs. Although the present 
of dark matter particles has been confirmed, the nature of them are still unknown. 
There are many theories of dark matter particles and the frequently researched one is the weakly 
interacting massive particles (WIMPs). According to the theory, WIMPs can 
annihilate into standard model particles such as photons, electrons, 
positrons and so on~\cite{Jungman:1995df,Bertone:2004pz}. Due to the basic quality of UCMHs, the annihilation rate 
of dark matter particles within them is very strong and these objects 
are the potential high energy astrophysical sources~\cite{Scott:2009tu}. By the research 
on the gamma-ray flux from the UCMHs due to the dark 
matter annihilation one can get the constraints on the cosmological abundance of them~\cite{Josan:2010vn,Bringmann:2011ut}. 
Further, because the formation of UCMHs is related to the primordial density 
perturbations of small scales, so the limits on their abundance can be 
converted to the limits on the PPS of small scales. 
In the Refs.~\citen{Josan:2010vn,Bringmann:2011ut}, the authors studied the 
gamma-ray flux from the UCMHs due to the 
dark matter annihilation and got the 
constraints on the PPS through comparing with the Fermi observations. They found that the strongest limit 
is about 5 orders stronger than that of PBHs. In addition to the gamma-ray flux, according to the dark matter theory, the 
neutrinos are usually produced going with the gamma-ray photons during the dark matter 
annihilation especially for the lepton channels. In the Ref.~\citen{Yang:2013dsa}, we studied the neutrino flux from the UMCHs 
due to the dark matter annihilation and investigated the limits on the PPS of small scales. 
We found that the strongest limit is about 5 orders stronger than that of PBHs. 

Beside annihilation, in some models dark matter 
particles are unstable and can decay into the standard model particles. This case is also very 
interesting in the indirect detection of dark matter~\cite{ypf,1475-7516-2009-02-021,1475-7516-2009-01-043}. 
In previous work, we studied the gamma-ray 
flux from the UCMHs due to the dark matter decay and got the $2\sigma$ upper  
limits on the PPS~\cite{EPL}. Because the decay rate is in proportion 
to the number density, so the constraints on the PPS are weaker than the 
annihilation case, but they are still about 4 orders stronger than the cases of PBHs. 
Recently, the high energy neutrino events are observed by the 
IceCube and dark matter decay is attracted much more interesting
~\cite{Esmaili:2013gha,Bhattacharya:2014vwa}. In the Ref.~\citen{raa}, the authors studied the 
neutrino flux from the UCMHs due to the gravitino decay. 
In this paper, we extend the analysis of that work and get the constraints on the PPS of small scales. 
  
This paper is organized as follows. In Sec.2, the main characters of UCMHs 
are introduced. In Sec.3I, the neutrino flux from UCMHs due to 
dark matter decay are calculated, the constraints on the mass fraction 
of UCMHs are given in Sec. 4. In Sec. 5, we get the limits on the PPS 
of primordial density perturbations. Finally, the conclusions are shown in Sec. 6.

\section{The Density Profile of UCMHs}
\label{density}
According to the structure formation theory, the density perturbations in the earlier epoch 
with the amplitude $\delta\rho/\rho \sim 10^{-5}$ can form the present cosmic structures. But if the amplitude of the density 
perturbations is larger than 0.3(or 0.7) then the primordial black holes 
(PBHs) are formed~\cite{Green:1997sz}. 
Recently, Ricotti and Gould proposed that if the density perturbations in early epoch 
were between $0.003<\delta \rho/\rho < 0.3$ one new kind of dark matter 
structures named ultracompact dark matter minihalos (UCMHs) could be formed
~\cite{Ricotti:2009bs}. Because the amplitude is 
not so large, so the formation probability of UCMHs 
is larger than that of PBHs. 
After the formation of UCMHs, dark matter particles and baryons are 
attracted through the radial infall. 
One dimension simulation indicates that 
the density profile of UCMHs is in the form as
~\cite{Ricotti:2009bs}

\begin{equation}
    \rho(r,z)=\frac{3f_{\chi}M_\mathrm{UCMHs}(z)}{16\pi R(z)^{\frac{3}{4}}r^{\frac{9}{4}}},
\label{eq:rho}
\end{equation}
where $M_{\mathrm{UCMHs}}(z)$ 
is the mass of UCMHs at redshift $z$, 
$R(z)=0.019(\frac{1000}{z+1})(\frac{M_\mathrm{UCMHs}(z)}{M_{\odot}})^{\frac{1}{3}}\mathrm{pc}$ 
is the radius of UCMHs at redshift $z$ and $f_{\chi}=\frac{\Omega_\mathrm{CDM}}
{\Omega_\mathrm{CDM}+\Omega_\mathrm{b}}=0.845$~\cite{planck}. 
From the Eq.~(\ref{eq:rho}) it can 
be seen that the density profile is in proportion to 
$r^{-2.25}$ and it is steeper than the Navarro-Frenk-White (NFW) profile ($\rho_{\mathrm{{NFW}}}(r) \sim 
r^{-1}$ for $r \to 0$) which has been used usually for the standard dark matter halos~\cite{nfw}. 
In fact, the center density of UCMHs is not infinitely large and it is usually 
affected by many effects. The main 
effect is the angular momentum of dark matter particles during being attracted through the radial infall~\cite{Bringmann:2011ut}. 
Following the Ref.~\citen{Bringmann:2011ut}, we set the minimal radius 
as 

\begin{eqnarray}
 r_{\mathrm{min}} = 3 \times 10^{-7}\mathrm{R_{UCMHs,z=10}}\left(
\frac{M^0_\mathrm{UCMHs}}{\mathrm{M_\odot}}\right)^{-0.06},
\end{eqnarray} 
For the radius $r<r_{\mathrm{min}}$, we assume that the density is constant, 
$\rho(r)_{r<r_{\mathrm{min}}}=\rho(r_{\mathrm{min}})$. For the other effects which can affect the 
center density of UCMHs one can 
refer to the Refs.~\citen{Bringmann:2011ut,Berezinsky:2013fxa,Berezinsky:2014wya}.

\section{The Neutrino Flux from UCMHs Due to the Dark Matter Decay}

Dark matter as the main component of the Universe has been confirmed by many
observations. But the nature of them are still unknown. There 
are many dark matter models now and some models show that the dark matter particles 
can decay into standard model particles~\cite{Bertone:2004pz}. 
The productions of dark matter decay can be photons, electrons, positrons or 
neutrinos~\cite{Allahverdi:2011sx,Cirelli:2012ut,Zant:2009sv,ypf,fenglei}. As mentioned in 
the section \ref{density}, because the center density profile of UCMHs is very steep, so 
the dark matter annihilation rate is very larger in there. 
In previous works~\cite{Scott:2009tu,Josan:2010vn,Bringmann:2011ut} the 
authors studied the gamma-ray flux from the UCMHs due to the dark matter annihilation. 
They found that the gamma-ray flux would achieve the threshold value of detectors such as 
EGRET or Fermi. 
In addition to the gamma-ray flux, according to the dark matter theory the neutrinos 
are usually produced accompanying the gamma-ray photons 
especially for the lepton channels. 
In previous work~\cite{Yang:2013dsa}, we investigated 
the neutrino flux from the UCMHs due to the dark matter annihilation. 
We found that the neutrino flux can excess the background neutrino flux which are mainly 
from the interaction between the cosmic ray and atoms in the atmosphere. 
Although the dark matter annihilation in the UMCHs is very interesting, 
if the dark matter particles are not annihilated the decay is another very important issue. 
So in this paper, we consider 
the neutrino flux from the UCMHs due to dark matter decay. We 
consider two popular dark matter models, the weakly interacting massive 
particles (WIMPs) and the gravitino. 
The gravitino is the lightest supersymmetric particle and 
they can decay into standard model particles in the presence of R-parity 
breaking~\cite{gravitino}. 
The decay channels considered here 
are $W^+W^-,b\bar b,\tau^+\tau^-,\mu^+\mu^-$ for the WIMPs. For the gravitino 
decay, there are mainly two-body and three-body decay channels. In this work, we 
mainly consider the three-body decay channel which has been used to explain the positrons excess~\cite{Bajc:2010qj}. 
We use the public code DarkSUSY~\cite{darksusy} to calculate the energy 
spectrum of neutrino for the WIMPs decay. For the gravitino decay, we use the forms given 
in the Ref.~\citen{Erkoca:2010vk}.

For the neutrino ($\nu_{\mu}$) detection, the main way is to detect the muons 
($\mu$) which are produced through the charged current interaction of neutrinos with the 
medium during propagation. There are two typical types of signal events. 
One is the upward events that the muons are produced 
out of the detection and another is the contained events that the muons are 
produced in the detection. In previous work, we considered these two 
cases for dark matter annihilation and found that the final limits 
on the PPS are better for upward events~\cite{Yang:2013dsa}. So 
in this work, we consider this case. The muon flux for upward events 
can be written as~\cite{erkoca,yq_1} 

\begin{eqnarray}
    \frac{d\phi_{\mu}}{dE_{\mu}}=\int^{m_{\chi}}_{E_{\mu}}dE_{\nu} \frac{d\phi_{\nu}}{dE_{\nu}}
    \frac{N_{A} \rho}{2}&&\left(\frac{d\sigma^{P}_{\nu}(E_{\nu},E_{\mu})}{dE_{\mu}}+(p\rightarrow n)\right)\nonumber \\
    &&\times R(E_{\mu})+(\nu\rightarrow\overline{\nu}),
\end{eqnarray}
where $R(E_{\mu})$ is the range which muons 
can propagate in matter until their energy is below the threshold of the detector 
$E_{\mu}^{th}$ and it is in the form of $R(E_{\mu})=\frac{1}{\beta\rho}
\ln(\frac{\alpha+\beta E_{\mu}}{\alpha+\beta E_{\mu}^{th}})$~\cite{neutrino_r_mu}, 
where $\alpha = 2.0 \times 10^{-6} \mathrm{TeV cm^2 g^{-1}}$ 
corresponds to the ionization energy loss and 
$\beta = 4.2 \times 10^{-6} \mathrm{cm^2 g^{-1}}$ 
accounts for the bremsstrahlung pair production and photonuclear interactions. 
$N_{A}=6.022\times10^{23}$ is Avogadro's number. $\rho$ is the 
density of medium and it is $0.918 \mathrm{g cm^{-3}}$ for ice. 
$d\sigma^{p,n}_{\nu,\overline{\nu}}/dE_{\mu}$ are the weak 
scattering charged-current cross sections for neutrino and antineutrino 
scattered off protons and neutrons~\cite{Covi:2009xn}. 
$d\phi_{\nu}/dE_{\nu}$ is the differential 
flux of neutrinos from UCMHs due to dark matter decay, 

\begin{equation}
    \frac{d\phi_{\nu}}{dE_{\nu}}= \frac{\Gamma}{m_{\chi} d^2}
\frac{dN_{\nu}}{dE_{\nu}}\left(\int^{\mathrm{r_{min}}}_{0} + 
\int^{\mathrm{R_{UCMH}}}_{\mathrm{r_{min}}}\right)\rho(r)r^2dr,
\label{eq:eq4}
\end{equation}
where $dN_{\nu}/dE_{\nu}$ is the energy spectrum of neutrino, 
$m_{\chi}$ and $\Gamma$ are the dark matter mass and decay rate, 
$d$ is the distance of UCMHs from the Earth.

The main background of neutrino detection is the atmosphere neutrinos (ATM) 
which are produced through the interaction between the cosmic ray and the 
atoms in the atmosphere. These neutrinos has been observed by the detectors such as IceCube~\cite{icecube}. 
The flux of ATM can be written as 
(in units of $\mathrm{Gev^{-1}km^{-2}yr^{-1}sr^{-1}}$) \cite{Erkoca:2010vk}

\begin{eqnarray}
   \left(\frac{d\phi_{\nu}}{dE_{\nu}d\Omega}\right)_\mathrm{ATM}&&=
N_{0}E_{\nu}^{-\gamma-1}\times \nonumber \\
&&\left(\frac{a\mathrm{ln}(1+bE_\nu)}{1+bE_{\nu}}+\frac{c\mathrm{ln}(1+eE_\nu)}{1+eE_{\nu}}\right),
\end{eqnarray}
where $a = 0.018, b= 0.024, c = 0.0069, e = 0.00139$, $\gamma = 1.74$ and 
$N_0 = 1.95(1.35) \times 10^{17}$ for $\nu(\bar\nu)$. 
In this work, we set the dark matter mass as $m_{\chi} 
= 1 \mathrm{TeV}$ and $10 \mathrm{TeV}$. Another important parameter is the decay rate and 
it has been 
constrained by many observations~\cite{Zhang:2007zzh,Huang:2011dq,DeLopeAmigo:2009dc}. 
In this work, we set the decay rate as 
$\Gamma = 10^{-26} s^{-1}$ and the final results can be applied easily for other values. 

\section{The Limits on the Mass Fraction of UCMHs}

After formation of the UCMHs, one of the important questions is the mass fraction of them in the Universe. 
In Refs.~\citen{Josan:2010vn,Bringmann:2011ut}, by 
researching the gamma-ray flux from the UCMHs due to the dark matter 
annihilation, the authors found the $2\sigma$ upper limit on the fraction of UCMHs 
is $f_\mathrm{{UCMHs,Anni.,\gamma}} \sim 10^{-7}$. We studied the gamma-ray flux 
for the dark matter decay~\cite{EPL}. Because the decay rate is in proportion to the 
number density of dark matter particles, so the limit is weaker than 
annihilation case, $f_\mathrm{{UCMHs,Dec.,\gamma}} \sim 10^{-5}$ ($2\sigma$ upper limit). In the Ref.~\citen{raa}, 
the authors researched the neutrino flux from UCMHs due to the 
gravitino decay and found that the $2\sigma$ upper limit is $f_\mathrm{{UCMHs,Dec.,\nu}} \sim 10^{-3}$.  
Although the limit 
is weak, if the dark matter particles are not annihilated the decay is very important. 

Following the Ref.~\citen{Bringmann:2011ut}, the fraction of 
UCMHs for non-observation of neutrino signals from UCMHs due to dark matter decay can be written as
\footnote{More general form can be found in the Ref.~\citen{Shandera:2012ke}(eq.(A2)). 
The difference of final results deduced by these two forms can be neglected safely for this work.} 

\begin{equation}
   f_\mathrm{UCMHs} = \frac{f_{\chi}M_\mathrm{{UCMH}}}{M_\mathrm{{MW}}}
\frac{\mathrm{log}(1-y/x)}{\mathrm{log}(1-M_{d<d_\mathrm{{obs}}}/M_\mathrm{{MW}})},
\label{eq:f}
\end{equation}
where $y$ and $x$ are the confidence level corresponding
to $f_\mathrm{{UCMHs}}$ and detector, respectively. Because for 
neutrino detection the ATM is the main background, so in this work we set the 
ATM as the non-detection upper limits. We times the number of ATM 
with 1.8 as the 
5$\sigma$ upper limits ($\sigma_\mathrm{{total,ATM}} = 16\%$~\cite{Abbasi:2010qv}). 
$M_\mathrm{{r<d,MW}}$ is 
the mass of dark matter halo within the radius $r<d$.

\begin{figure}
\begin{center}
\epsfig{file=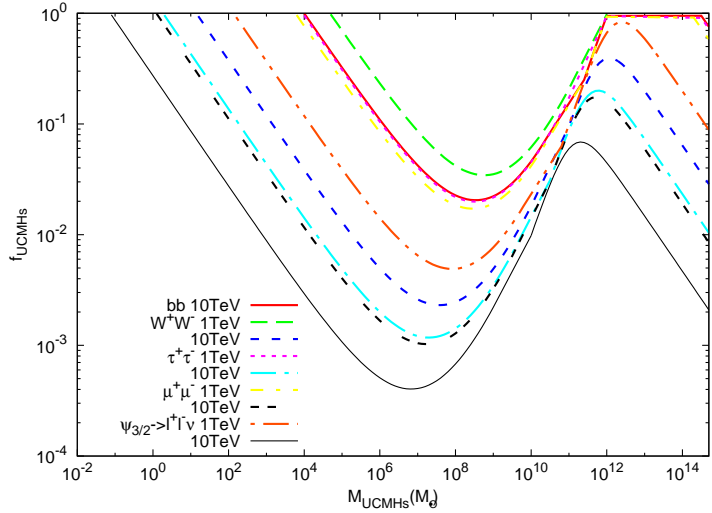,width=0.8\textwidth}
\caption{The $2\sigma$ upper limits on the mass fraction of the UCMHs for different decay channels, 
$\chi \to W^+W^-,b\bar b,\tau^+\tau^-,\mu^+\mu^-$ and 
$\psi_\frac{3}{2} \to l^{+}l^{-}\nu$ for $l=\mu$. The dark 
matter mass is 1TeV and 10TeV, the decay rate is $\Gamma = 10^{-26} s^{-1}$, 
the exposure time of detector is 10 years.}
\label{fig:fraction}
\end{center}
\end{figure}

Following previous works~\cite{Yang:2013dsa,raa}, we calculate the neutrino 
number from a UCMH due to dark matter decay with some confidence level 
(e.g. 2$\sigma$) for some exposure times (e.g. 10 years) using the formula~\cite{Bergstrom:1997tp}

\begin{equation}
    T_\mathrm{{obs}}=\sigma^{2}\frac{N_\mathrm{{ATM}}+N_\mathrm{{UCMHs}}}{N_\mathrm{{UCMHs}}^{2}},
\end{equation}
where $T_\mathrm{{obs}}$ is the exposure time, $N_\mathrm{{UCMHs}}$ is the neutrino 
number from a UCMH due to 
dark matter decay and it can be obtained by the integration
\begin{equation}
    N_\mathrm{UCMHs}=\int^{E_\mathrm{max}}_{E_{\mu}^{th}}
\frac{d\phi_{\mu}}{dE_{\mu}}A_\mathrm{eff}(E_{\mu},\theta)dE_{\mu},
\end{equation}
where $A_\mathrm{eff}$ is the effective area of detection, it 
is a function of energy and zenith angle~\cite{Erkoca:2009by}. For a fixed exposure time, e.g. 10 years, 
the distance of UCMH can be obtained from the Eq.~\ref{eq:eq4}. Then 
the upper limits on the mass fraction of UCMHs for 2$\sigma$ confidence level can be obtained using the Eq.~\ref{eq:f}. 
The results are given in Fig.~\ref{fig:fraction} where different decay channels mentioned 
in previous section are shown. From these plots it can be found 
that for dark matter decay, the strongest limit 
on the fraction of UCMHs is from the lepton channels and large dark matter mass, 
the strongest $2\sigma$ upper limit is $f_\mathrm{UCMHs,Dec.,\nu} \sim 4 \times 10^{-4}$ for 
gravitino decay with UCMH mass $M_\mathrm{{UCMHs}} \sim 10^{7} M_{\odot}$. 

In the Ref.~\citen{raa}, the authors also studied the limits on the mass faction 
of UCMHs for gravitino decay. The processes of calculations in that paper are 
slightly different from this work. For example, for the effect area of detector 
$A_\mathrm{eff}$, they assumed a constant value $A_\mathrm{eff} = 1 \mathrm{km^2}$. 
In this work, we used a general form which depends on the energy and zenith angel. 
More over, for the definition of mass fraction of UCMHs they used a simple one following 
the Ref.~\citen{Josan:2010vn}. So the final limits on the mass fraction of UCMHs 
are different from that of this work for the same decay channel and dark matter mass. 

\section{Constraints on the Primordial Power Spectrum}

As mentioned in above section many inflation models predict that the PPS 
is scale invariant over a wide scale ranges and some 
other models show that the PPS is scale dependent~\cite{lidsey,Joy:2007na}. 
Therefore, the limits on the PPS is very important for checking different 
inflation models. The main limits at present focus on the large scales 
which are mainly from the observations on the CMB, Large-scale structures and Lyman-$\alpha$ forest. 
For small scales, the main limits are from the research on PBHs
\footnote{The study on the dissipation of acoustic waves can also be used to constrain the PPS~\cite{Nakama:2014vla}}, 
but these constraints are very weak. 
The research on UCMHs provides another better way to study the PPS of small scales. 
More detailed calculations of limits on the PPS are given in Ref.~\citen{Bringmann:2011ut} and 
in this paper we only give the main points.

If the initial perturbations are Gaussian the present mass fraction of UCMHs can 
be written as~\cite{Bringmann:2011ut}

\begin{eqnarray}
\mathrm{\Omega_{UCMHs}} = &&\mathrm{\frac{\Omega_{DM}}{\sqrt{2\pi}\sigma_{H}(R)}
\frac{M_{UCMHs,z=0}}{M_{UCMHs,z_{eq}}}} \nonumber \\&& 
\times \int^{\sigma_{\mathrm{max}}}_{\sigma_\mathrm{{min}}}
\mathrm{exp}\left(-\frac{\sigma^2_{H}(R)}{2\sigma^2_{H}(R)}\right)d\sigma_{H}(R),
\end{eqnarray}
where $\sigma_{\mathrm{max}}$ and $\sigma_{\mathrm{min}}$ are the maximal 
and minimal values of density perturbations required for the formation 
of UCMHs. These values are the function of redshift. Following the Ref.~\citen{Bringmann:2011ut}
we use the values corresponding to the redshift, $z=1000$, at which the UCMHs are 
formed. The PPS ($\mathcal{P}_{\mathcal{R}}$) is related to the $\sigma_{H}(R)$ as 

\begin{equation}
\sigma^2_{H}(R) = \frac{1}{9}\int^{\infty}_{0}x^3W^2(x)\mathcal{P}_{\mathcal{R}}
(x/R)T^2(x/\sqrt{3})dx,
\end{equation} 
where $W(x) = 3x^{-3}(sinx-xcosx)$ is the Fourier transform of the top-hat 
windows function with $x \equiv kR$. $T$ is the transfer function of 
the evolution of perturbations. For more detailed discussions one can see the 
appendixes in Refs. \citen{Bringmann:2011ut,Li:2012qha}. 

\begin{figure}
\begin{center}
\epsfig{file=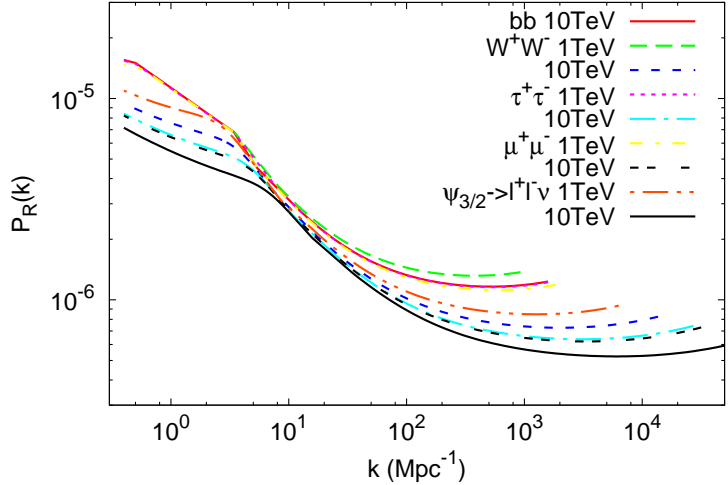,width=0.8\textwidth}
\caption{The $2\sigma$ upper limits on the primordial power spectrum of small scales 
for different decay channels and dark matter mass. The related values of parameters 
are the same as the Fig.~\ref{fig:fraction}.}
\label{fig:fig2}
\end{center}
\end{figure}

The constraints 
on the PPS are plotted in the 
Fig.~\ref{fig:fig2}. From these plots it can be seen that the strongest $2\sigma$ upper 
limit is $\mathcal{P}_{\mathcal{R}} \sim 3 \times 10^{-7}$ for $k \sim 5 \times 10^{3} \mathrm{Mpc^{-1}}$. 
This limit is comparable with the results of Ref.~\citen{Bringmann:2011ut}. 
In that work, the authors investigated the gamma-ray flux from UCMHs 
due to the dark matter annihilation and got the limits on the PPS for the $b \bar b$ 
channel and for the dark matter mass $m_{\chi} = 1 \mathrm{TeV}$. But in this work, 
one can find that the limits for the $b \bar b$ channel are weaker than 
the lepton channels. 

For these 
results one should notice that they depend on the character of dark matter 
particles and other aspects such as the density profile of dark matter 
halo of Milky Way. In previous paper~\cite{EPL}, we researched the dependence of the constraints 
on the different density profiles of dark matter halo and dark matter decay rate. From 
that results one can conclude that the constraints are stronger for the 
NFW density profile or the large decay rate. Another important factor 
is the dark matter particle mass. From the results of this work and the 
Refs.~\citen{Yang:2013dsa,EPL} it can be seen that the limits on the PPS 
are stronger for the larger dark matter mass.

\section{Conclusions}
The research on the PPS is very important for checking different inflation models. 
At present, the main constraints on the PPS focus on the large scales. For small scales 
the main limits are from the study on the PBHs. 
Because the formation of UCMHs is related to the PPS of small scales, 
so the constraints on the mass fraction of UMCHs can be converted to 
the limits on the PPS.
In this work, we considered the neutrino flux from the UCMHs 
due to the dark matter decay. We found that the $2\sigma$ upper limit is 
$f_\mathrm{UCMHs} \sim 4 \times 10^{-4}$ for $M_\mathrm{{UCMHs}} \sim 10^{7} M_{\odot}$. 
For the limits on the PPS of small scales, the limits 
are $\mathcal{P}_{\mathcal{R}} \lesssim 3 \times 10^{-7}$ for $k \sim 5 \times 10^{3} \mathrm{Mpc^{-1}}$ 
with $2\sigma$ confidence level.
These constraints are comparable with that of Ref.~\citen{Bringmann:2011ut,Yang:2013dsa}, 
but the corresponding scales are different.

\section{Acknowledgments} Yupeng Yang is supported by the 
National Science Foundation for Post-doctoral Scientists of China (Grant No.2013M540434) 
and National Science Foundation of China (Grant No.11347148 and No.U1404114). 

%\bibliographystyle{ws-ijmpa.bst}
%\bibliography{bibtex}

\end{document}